\documentclass[aps,prb,twocolumn,superscriptaddress,showpacs]{revtex4}

\usepackage{graphicx}
\usepackage{natbib}

\newcommand{\gtappr}{{{\lower4pt\hbox{$>$} } \atop \widetilde{ \ \ \ }}}
\newcommand{\ltappr}{{{\lower4pt\hbox{$<$} } \atop \widetilde{ \ \ \ }}}

\newcommand{\beq}{\begin{equation}}
\newcommand{\eeq}{\end{equation}}

\newcommand{\ud}{\mathrm{d}}

\newcommand{\ie}{\textit{i.e. }}

\newcommand{\ybal}{$\beta$-YbAlB$_4\,$}
\newcommand{\aybal}{$\alpha$-YbAlB$_4\,$}

\newcommand{\yrs}{YbRh$_2$Si$_2\,$}

\def\gsim{\buildrel {\textstyle >}\over {_\sim}}
\def\lsim{\buildrel {\textstyle <}\over {_\sim}}

\newsavebox{\fmbox}

\begin{document}


\title{Field evolution of quantum critical and heavy Fermi-liquid components in the magnetization of the mixed valence compound \ybal}


\author{Yosuke Matsumoto}
\email[]{matsumoto@issp.u-toyko.ac.jp}
\affiliation{Institute for Solid State Physics, University of Tokyo, Kashiwa, Chiba 277-8581, Japan}

\author{K. Kuga}
\affiliation{Institute for Solid State Physics, University of Tokyo, Kashiwa, Chiba 277-8581, Japan}

\author{Y. Karaki}
\affiliation{Institute for Solid State Physics, University of Tokyo, Kashiwa, Chiba 277-8581, Japan}
\affiliation{Faculty of Education, University of the Ryukyus, Nishihara, Okinawa 903-0213, Japan}

\author{Y. Shimura}
\affiliation{Institute for Solid State Physics, University of Tokyo, Kashiwa, Chiba 277-8581, Japan}

\author{T. Sakakibara}
\affiliation{Institute for Solid State Physics, University of Tokyo, Kashiwa, Chiba 277-8581, Japan}

\author{M. Tokunaga}
\affiliation{Institute for Solid State Physics, University of Tokyo, Kashiwa, Chiba 277-8581, Japan}

\author{K. Kindo}
\affiliation{Institute for Solid State Physics, University of Tokyo, Kashiwa, Chiba 277-8581, Japan}

\author{S. Nakatsuji}
\email[]{satoru@issp.u-tokyo.ac.jp}
\affiliation{Institute for Solid State Physics, University of Tokyo, Kashiwa, Chiba 277-8581, Japan}


\date{\today}

\begin{abstract}
We present the high-precision magnetization data of the 
valence fluctuating heavy fermion superconductor \ybal in a wide temperature range from 0.02 K to 320 K spanning four orders of magnitude. 
We made detailed analyses of 
the $T/B$ scaling of the magnetization, and firmly confirmed the unconventional zero-field quantum criticality (QC) without tuning. 
We examined other possible scaling relationship such as $T/(B-B_c)^{\delta}$ scaling, and confirmed that 
$\delta = 1$ provides the best quality of the fit with an upper bound on the critical magnetic field $\vert B_c \vert <0.2$~mT. 
We further discuss the heavy Fermi-liquid component of the magnetization after subtracting the QC component estimated based on the $T/B$ scaling. 
The temperature dependence of the heavy Fermi-liquid component is found very similar to the magnetization of the polymorph \aybal.
In addition, the heavy Fermi-liquid component is suppressed in the magnetic field above $\sim$ 5 T as in \aybal. 
This was also confirmed by the magnetization measurements up to $\sim 50$ T for both $\alpha$- and \ybal. 
Interestingly, the detailed analyses revealed that the only a part of $f$ electrons participates in the zero-field QC and the heavy fermion behavior. 
We also present a temperature - magnetic field phase diagram of \ybal to illustrate how the characteristic temperature and field scales evolves near the QC.

\end{abstract}

\pacs{71.27.+a, 71.28.+d, 74.40.Kb, 75.20.Hr, 75.30.Mb}

\maketitle


\section{Introduction}
Formation of novel quantum phases in the vicinity of a quantum critical point (QCP) 
has been studied extensively in condensed matter physics 
for a past few decades. 
Especially, in the heavy fermion intermetallic systems, a number of prototypical examples of novel phenomena, such as 
unconventional superconductivity and non-Fermi liquid (NFL) behavior, have been discovered 
in the vicinity of a magnetic QCP where the magnetic ordering temperature is suppressed to zero~\cite{Mathur98,Lohneysen07,gegenwart08}. 

So far, these studies of quantum criticality (QC) have been restricted mostly to the Kondo lattice systems with integer valence.  
On the other hand, the first Yb-based heavy fermion superconductor \ybal provides a unique example of a QC 
in the strongly mixed valence state \cite{nakatsuji08, KugaPRL, matsumoto-ZFQCP, ybal-valency}. 
Furthermore, the QC cannot be described by the standard theory for the spin-density-wave instability\cite{Hertz76, Moriya85, Millis93}.  
The diverging magnetic susceptibility along the $c$-axis exhibits $T/B$ scaling in the wide temperature ($T$) and magnetic field ($B$) region spanning 3 $\sim$ 4 orders of magnitude. 
This indicates that the QCP is located just at the zero magnetic field within the experimental resolution of  0.2~mT under ambient pressure \cite{matsumoto-ZFQCP}. 
The QC emerging without tuning any control parameter suggests a formation of an anomalous metallic phase. 

\ybal has the locally isostructral polymorph \aybal, which has, in contrast, a Fermi liquid (FL) ground state at zero field \cite{matsumoto-PRB84}. 
\aybal is also strongly mixed valent. The Yb valence estimated by a hard x-ray photoemission spectroscopy is +2.73 
for \aybal and +2.75 for \ybal at 20 K \cite{ybal-valency}.  
The valence fluctuation temperature scale was estimated to be $\sim 300$ K 
from X-ray adsorption measurements for the both systems \cite{matsuda-X-ray-JKPS62}.  
Correspondingly, peaks have been found in the magnetic part of the in-plane resistivity ($\rho_{ab}^m$) and the in-plane magnetic susceptibility 
in the same temperature range of $200 < T < 300$ K. 
Nevertheless, these two systems exhibit a heavy fermion (HF) state with Kondo lattice like localized moments far below the valence fluctuation scale \cite{matsumoto-ZFQCP}.  
This is quite unusual because Pauli paramagnetism is usually expected in the mixed valence compounds below the valence fluctuation temperature scale. 
The small temperature scale of $\sim 8$ K for the anomalous HF state may indicate that \aybal is also close to a QCP.  
The origin of the HF state is a key to understand the novel QC in \ybal, which challenges the conventional understanding of the QC 
based on the so called Doniach phase diagram.  

Here, we discuss the magnetization ($M$) in \ybal in the $T$ range from 0.02 K to 320 K spanning four orders of magnitude 
and see how it evolves with magnetic field.  
First, we provide a $T$-$B$ phase diagram in order to overview the various $T$ and $B$ scales for \ybal. 
Then, we present the temperature dependence of $-dM/dT$ in a wider $T$ region between
0.02 K and 320 K than the previous report \cite{matsumoto-ZFQCP}. 
To verify the zero field quantum criticality, we examine a possibility of other scaling such as $T/(B-B_{\rm c})^{\delta}$ scaling with $B_c\neq 0$ or $\delta\neq 1$, 
and confirm that the $T/B$ scaling reported in our previous work provides the best quality of the fitting. 

We will further discuss the heavy Fermi-liquid component of $M$ after subtracting the QC component estimated by using the $T/B$ scaling. 
The obtained heavy Fermi-liquid component exhibits the $T$ dependence quite similar to \aybal having a peak in $-dM/dT$ at $\sim 8$ K. 
Recently, it has been revealed that the HF state in \aybal is suppressed in the magnetic field above $\sim$ 5 T~\cite{matsumoto-SCES2013}.
Here we found that the heavy Fermi-liquid component of $M$ in \ybal also exhibits a field evolution quite similar to the one for \aybal, 
suggesting that the HF state in \ybal is also suppressed in the magnetic field above $\sim$ 5 T. 
This was also confirmed by the measurements of the magnetization curve up to $\sim 50$ T for both $\alpha$- and \ybal. 
In addition, it further indicates that only a part of $f$ electrons participates in the zero-field QC and exhibits the HF behavior.

Note that partial results presented here have already been discussed in Refs.~\cite{matsumoto-ZFQCP, matsumoto-PSSB247, matsumoto-SCES2010
}.  
The experimental details of this work can be found in Refs.~\cite{matsumoto-ZFQCP,matsumoto-PSSB247}. 
We also note that $M$ was measured with the resolution of $\sim 10^{-8}$ emu by 
the high precision SQUID magnetometer installed in a $^3$He-$^4$He dilution refrigerator at $T < 4$ K and $B < 0.05$ T. 
The residual magnetic field was as small as $\sim$ 1.1 $\mu$T. 
At $T < 4$ K and $B\geq  0.05$ T, it was measured with a resolution of $\sim 10^{-5}$ emu by a high precision Faraday magnetometer 
installed in a $^3$He-$^4$He dilution refrigerator\cite{sakakibara_Faraday}. 
We employed only high quality single crystals after carefully washed the surface of each crystal to remove possible impurities.
The residual resistivity ratio (RRR) of the samples used for the SQUID magnetometer measurements ($\sim$ 30 pieces, 0.82 mg) was higher than 200. 
For the Faraday magnetometer measurements, we used single crystals of 7.5 mg, 
whose typical RRR is as high as 140. 
The higher $T$ measurements above 2 K were done by a commercial SQUID magnetometer. 
High field magnetization curves up to $\sim 50$ T were measured using a
pulsed field facility at the ISSP of the University of Tokyo. 
Both of these two sets of measurements were done for both \ybal and \aybal 
for the samples with typical RRR of 140 and 50, respectively.

\section{Results and Discussion}
\subsection{$T$-$B$ phase diagram}
\begin{figure}[tb]
\begin{center}
\includegraphics[width = 7.7 cm,clip]{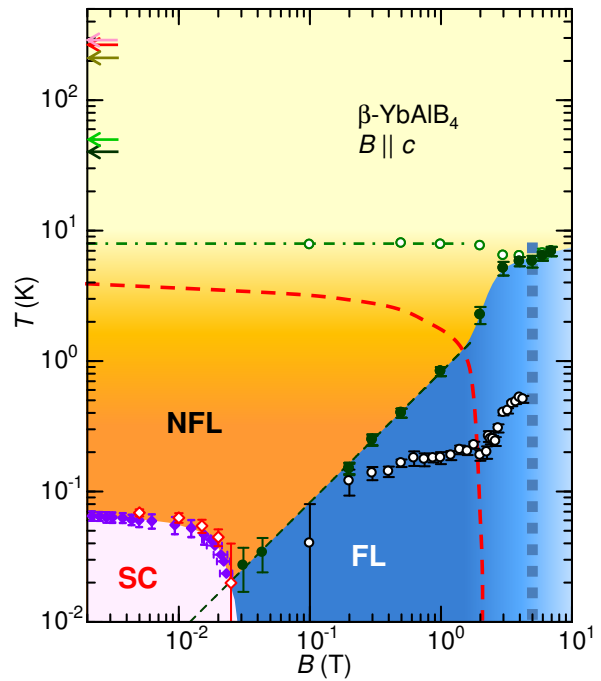}
\end{center}
\caption{
$T$-$B$ phase diagram of \ybal for the field applied along the $c$-axis. 
The valence fluctuation scale is shown as pink, red and dark yellow arrows at $T\sim$ 200 - 300 K 
which corresponds to $T$ scale obtained by X-ray adsorption\cite{matsuda-X-ray-JKPS62}, 
the coherence peak in $\rho _{ab}^m$\cite{nakatsuji08} and a peak in $\chi _{ab}$ at $B=$ 0.1, 7.0 T, respectively. 
The light and dark green arrow indicate the anomaly in the $T$ derivative of the in-plane resistivity $d\rho _{ab} /dT$ at $T\sim 50$ K\cite{matsumoto-PRB84} and 
the peak in the Hall coefficient $R_H$ at $\sim 40$ K\cite{eoin-PRL109}, respectively.  
These temperature scales correspond to the effective Kondo lattice temperature for the low $T$ HF behavior.
Green filled circles correspond to the peaks in $\-d\chi/dT$, which separates the QC and FL regions. 
Green open circles correspond to the peak temperature scale $T^*$ in $\Delta (-d\chi/dT)$ 
obtained after subtractiong QC components, which sets the onset of the heavy fermion state. 
Black open circles correspond to $T_{\rm FL}$ determined by $\rho _{ab}$ \cite{nakatsuji08}. 
Filled and open diamonds correspond to superconducting phase boundary 
determined by the SQUID ac susceptibility measurements\cite{matsumoto-PSSB247} and the resistivity measurements\cite{KugaPRL}, respectively. 
Inside the red broken line, the QC scaling is observed. 
The vertical blue broken line at $B\sim 5$ T indicates the crossover field above which the HF state is suppressed.   
For detail, see text.
 } 
\label{Fig_phase_diagram}
\end{figure}

In order to overview the characteristic $T$ and $B$ scales in \ybal, we first present a $T$-$B$ phase diagram for the field applied along the $c$-axis in Fig. \ref{Fig_phase_diagram}. 
To construct the phase diagram, 
we used all the data available so far including those we will present and discuss later in this paper.

Let us begin with the high $T$ region of the phase diagram. 
A pink arrow at $T\sim 290$ K indicates the valence fluctuation scale estimated by the X-ray adsorption measurements\cite{matsuda-X-ray-JKPS62}. 
The coherence peak observed in the in-plane resistivity $\rho _{ab}$ at $T\sim 270$ K (red arrow) and 
a peak in $\chi _{ab}$ found at $T\sim 210$ K for $B=0.1$, 7.0 T (dark yellow arrow) also correspond to the valence fluctuation scale. 
On cooling, another scale $\sim50$ K emerges as the effective Kondo temperature for the low $T$ HF behavior. 
For example, $M$ starts to increase below $T\sim 40$ K signaling the onset of the HF behavior, as we will discuss. 
At the same $T$-range, the Hall coefficient $R_H$ exhibits a pronounced peak indicating a coherence developing among $f$ electrons\cite{eoin-PRL109}. 
In addition, the $T$ derivative of the in-plane resistivity $d\rho _{ab} /dT$ shows a shoulder-like anomaly at $T\sim 50$ K\cite{matsumoto-PRB84}. 

Far below the valence fluctuation scale, we find 
another $T$ scale for the HF behavior at $T^*\sim 8$ K, typically in the susceptibility $\chi = M/B$ along the $c$-axis.  
As we will discuss later in detail, 
$\chi (T)$ along the $c$-axis consists of two components, \ie, the QC and the heavy Fermi-liquid components.  
$T^*$ is defined as a peak temperature of the heavy Fermi-liquid component of $-d\chi/dT$ in \ybal which is represented by $\Delta(-d\chi/dT)$.  
For \aybal, $T^*$ denotes the peak in $-d\chi/dT$. 
Below $T^*$, \ybal indicates the unconventional zero-field QC, while \aybal exhibits the FL properties at $B=0$. 

The $T/B$ scaling of $M$ was observed in the regime below $B\leq  2$ T and $T\leq  3$-4 K 
which is a region inside the red dashed line in Fig. \ref{Fig_phase_diagram}.  
Inside the region, the QC component exceeds the heavy Fermi-liquid component. 
The NFL-FL crossover is defined by the peak in $-dM/dT$ (or $-d\chi /dT$) and is observed on the line where $T/B\sim 0.8$ K/T. 
This linear field dependence of the peak corresponds to the Zeeman energy scale of an effective moment $\sim1.94$  $\mu _{\rm B}$\cite{matsumoto-ZFQCP,matsumoto-PSSB247}. 
The peak in $-dM/dT$ merges into the one for $\Delta(-d\chi/dT)$ at $B\geq 2$ T where the heavy Fermi-liquid component is dominant. 
The zero-field QCP is masked by the HF superconductivity observed at $T\leq  0.08$ K and $B\leq  0.03$ T along the $c$-axis. 
Note that the small $T_{\rm c}$ enables us to reveal the zero-field QC behaviors in detail. 
$\rho _{ab}$ exhibits $T^2$ behavior below $T_{\rm FL}$ which is much lower than the NFL-FL crossover temperatures. 
The HF behavior is suppressed above $B^*\sim 5$ T, 
a characteristic field scale determined by $M$ (the vertical thick dashed line in dark blue in Fig. \ref{Fig_phase_diagram}). 
$T_{\rm FL}$ shows a sudden increase at 2.3 T suggesting the suppression of the HF behavior 
at a slightly smaller field than $B^*\sim 5$ T. We will further discuss this in detail later. 

Thus, both \ybal and \aybal have the three contrasting temperature scales, \ie, 
the valence fluctuation scale of 200 - 300 K, the effective Kondo temperature $\sim 50$ K, and the HF $T$ scale of $T^*\sim 8$ K. 

\subsection{Overview of the magnetization data}

\begin{figure}[tb]
\begin{center}
\includegraphics[width = 8.3 cm,clip]{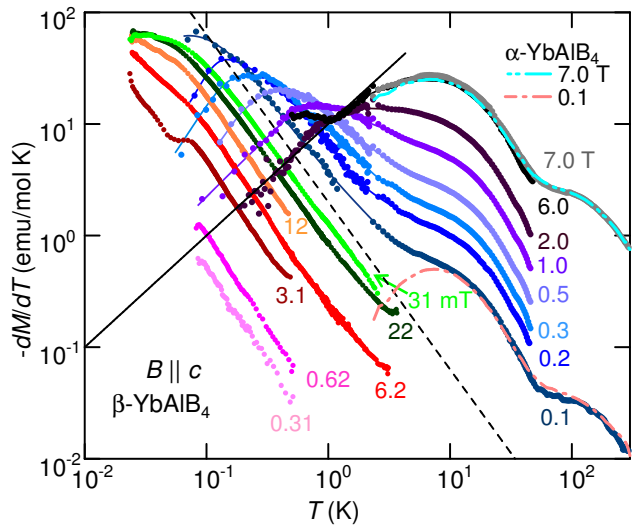}
\end{center}
\caption{$-dM/dT$ versus $T$ for selected fields along the $c$-axis on a logarithmic scale 
for both $\beta$- (solid circles) and \aybal  (long dashed dotted line for $B=0.1$ T and long dashed double-dotted line for $B=7.0$ T). 
The sudden downturn below 0.1 K in the low field data corresponds to the onset of superconductivity. 
 }
\label{Fig_MoverH-dMdT}
\end{figure}

Next, we discuss the magnetization ($M$) data measured over four decades of $T$ (0.02 K to 320 K) and $B$ (0.3 mT to 7 T) in detail. 
The divergent susceptibility $\chi = M/B$ along the $c$-axis is 
one of the most remarkable features 
of the QC in \ybal\cite{matsumoto-ZFQCP}. 
In order to discuss the divergent behavior of $\chi$ as a function of $T$ and $B$, we show 
 $-dM/dT$ versus $T$ for selected fields along the $c$-axis on a logarithmic scale in Fig. \ref{Fig_MoverH-dMdT} for 
both $\alpha$- and \ybal. 
As is clearly seen in the figure, 
below $T\sim$ 3 K, $-dM/dT$ for the $\beta$ phase shows a power law behavior with $T^{-1.5}$ dependence (broken line), 
indicating the divergence of the susceptibility $\chi\sim T^{-1/2}$ in the non-Fermi liquid region at $T/B \gtrsim 1$ K/T. 
On the other hand, a $T$-linear behavior (solid line) is observed at $T/B \lesssim 1$ K/T, which is expected for a FL state. 
By applying magnetic field, the quantum critical component is suppressed and finally masked by another component which appears at $T^* \sim 8$ K. 
At $B = 7.0$ T, it overlaps on top of the one for \aybal at least down to $T \sim 2$ K. 
This crossover at  $T^* \sim 8$ K is commonly seen in both $\beta$- and $\alpha$-YbAlB$_4$ and  gives characteristic $T$-scale of the Kondo lattice behavior. 
The low $T$ magnetization of \ybal is well expressed by a sum of the QC component $M_c/B\propto T^{-0.5}$ 
and $T$-independent constant term $\chi_0 (= M_0/B) = 0.017$ (emu/mol) 
which is close to the zero-$T$ susceptibility of the non-critical \aybal 
and may originate from a constant Van Vleck component to the susceptibility and/or from Pauli susceptibility of the non-critical parts of the Fermi surfaces. 
This heavy Fermi-liquid component is extracted after subtracting the QC component. 
We will further discuss the detail of this component later. 

\begin{figure*}[tb]
\begin{center}
\includegraphics[width =15.5 cm,clip]{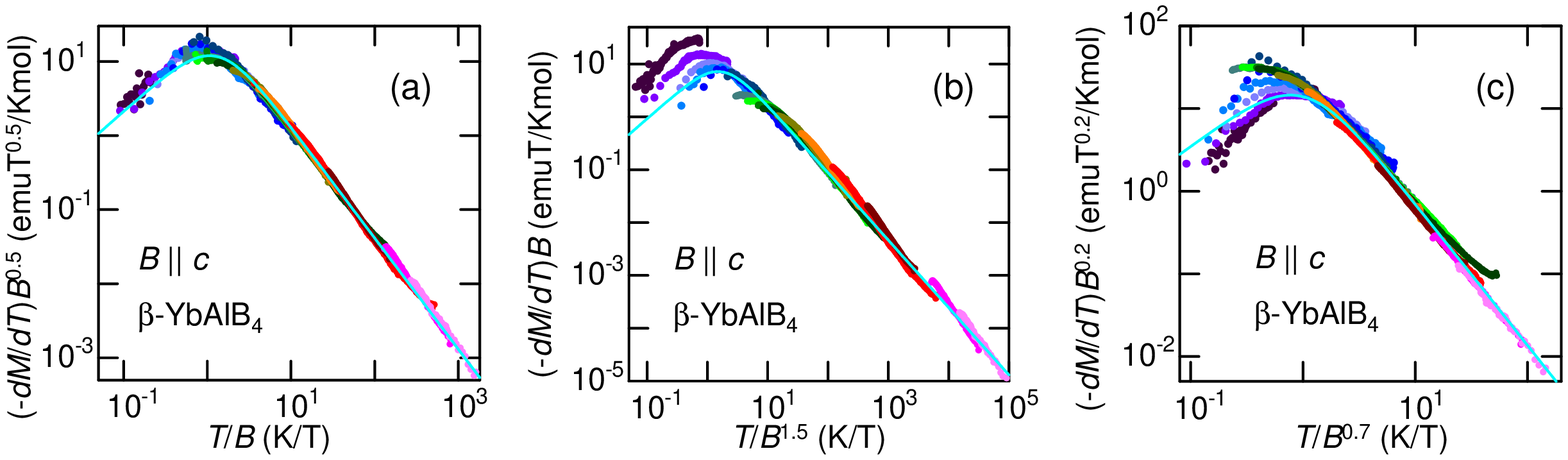}
\end{center}
\caption{Scaling plots $(-\ud M/\ud T) B^{\eta}$ vs $T/B^{\delta}$ with (a) $(\delta , \eta) = (1.0, 0.5)$, (b) $(\delta , \eta) = (1.5, 1.0)$, (c) $(\delta , \eta) = (0.7, 0.2)$ 
for the data shown in Figure \ref{Fig_MoverH-dMdT}(a) inset. 
The light blue lines are the fit by $\phi(x) = \Lambda x(A+x^2)^{\frac{\alpha}{2}-2}$ (see text). 
 }
\label{Fig_scaling}
\end{figure*}

\subsection{$T/B$ scaling and the evaluation of its quality}
As we already reported in the previous paper~\cite{matsumoto-ZFQCP}, 
if we plot $(-dM/dT)B^{0.5}$ vs. $T/B$ for \ybal, 
all the data in the region of $T \lesssim 3$~K and $B \lesssim 2$~T enclosed by the red broken line in Fig. \ref{Fig_phase_diagram} 
collapse on a single curve (Fig. \ref{Fig_scaling} (a)). 
This indicates a scaling relation of 
\beq 
-\frac{\ud M}{\ud T} = B^{\alpha -2}\phi \left(\frac{T}{B} \right),  
\label{dMdT} 
\eeq 
with $\alpha = 3/2$. 
The reason why we use $-dM/dT$ rather than $M_{\rm c}=M - M_0$ is that $-dM/dT$ is free from 
the ambiguity in the estimate of $\chi_0$ and its field evolution. 
In fact, the scaling obtained for $-dM/dT$ looks better than that of $M_{\rm c}$, \ie, $M_c/B^{0.5}$ vs. $T/B$ ~\cite{matsumoto-ZFQCP, matsumoto-SCES11}.
This empirical scaling relation 
implies that close to the QCP, 
\ybal has no intrinsic energy scale and the  
ratio $T/B$ determines the physical properties. 
The peak of the scaling curve is located at $T/B \sim 0.8$ K/T, defining 
the thermodynamic boundary between the FL and NFL regions in the $T$-$B$ phase diagram (Fig. \ref{Fig_phase_diagram}). 
As already discussed in the previous report\cite{matsumoto-ZFQCP}, 
the scaling behavior can be well fitted to Eq. (\ref{dMdT}) with 
\beq 
\phi(x) = \Lambda x(A+x^2)^{\frac{\alpha}{2}-2},  
\label{scalingfunc} 
\eeq 
for $\alpha = 3/2$. 
This form was chosen to satisfy the limiting behaviors of the QC free energy $F_{\rm QC}= B^{\alpha} f(T/B)$ obtained after integrating the both parts of Eq.~(\ref{dMdT}).  
Here $f(x)$ is related to $\phi (x)$ by $\phi (x)=(\alpha-1)f'(x)-xf''(x)$. 
Indeed, this satisfies 
$f(x)\propto x^{\alpha}$ in the NFL regime ($x \gg 1$) and $f(x)\propto \mathrm{const} + x^2$ in 
the FL phase ($x\ll 1$). 

The $T/B$ scaling implies that the critical field $B_{\rm c}$ of the quantum
phase transition is located just at zero field because a finite value of $B_{\rm c}$ requires the scaling function 
$\phi(x)$ with a ratio $x=T/|B-B_c|$ rather than $x=T/B$. 
In the previous work, $B_{\rm c}$, which gives the best fitting of the experimental data to Eq.~(\ref{dMdT}),  
was estimated to be $-0.1 \pm 0.1$~mT by using Eq.~(\ref{scalingfunc}). 
This is comparable to the Earth's magnetic field ($\sim 0.05$~mT) 
and two orders of magnitude smaller than $\mu_0 H_{\rm c2} = 30$ mT.
This tiny value indicates zero-field quantum criticality without tuning in \ybal\cite{matsumoto-ZFQCP}.  

If we take into account 
a possibility of other scaling, such as $T/B^{\delta}$ scaling with $\delta \neq 1$ and a possibility of finite $B_{\rm c}$, 
the scaling relation in Eq.~(\ref{dMdT}) will be generalized as follows. 
\beq 
-\frac{\ud M}{\ud T} = (B-B_{\rm c})^{-\eta}\phi \left(\frac{T}{(B-B_{\rm c})^{\delta}} \right).
\label{scaling} 
\eeq 
The qualitative evaluation of these possible scaling relations can be made by plotting 
$-(\ud M/\ud T)(B-B_{\rm c})^{\eta}$ versus $T/(B-B_{\rm c})^{\delta}$ for the same data set for the region shown in Fig.~\ref{Fig_MoverH-dMdT}(a) inset
while changing $B_{\rm c}$, $\delta$ and $\eta$. 
The examples of the other plots with ($B_{\rm c}$, $\delta$, $\eta$) = (0, 1.5, 1.0) and (0, 0.7, 0.2) 
are shown in Fig. \ref{Fig_scaling} (b) and (c) respectively. 
Compared to the scaling plot in Fig. \ref{Fig_scaling} (a), the data in these two examples do not collapse on a single curve. 

In order to quantitatively evaluate the quality of the scaling, 
here we use the correlation $R$ obtained after fitting $\phi (x)$ with $x=T/(B-B_{\rm c})^{\delta}$
to the data for each set of the parameters. 
Here $R$ is defined as $R\equiv \sqrt{1-\chi ^2/{\rm DOF}}$ where $\chi ^2$ is a statistical one 
and DOF (degree of freedom) is a number of the data points (2227 points). 
$R$ takes its maximum value of 1 if the fitting quality is perfect. 
In order to obtain a reasonable fit in a log-log scale, 
we fit the data with the weight of $(1/\sigma _i)^2=(-(\ud M_i/\ud T)(B-B_c)^{\eta})^2$ 
for $i$-th data. 
Otherwise, the power law behavior observed at the large $x$ region cannot be fitted well due to its small value. 
We checked a parameter region of -0.5 mT $\le B_c \le $ 0.2 mT, 
0.5 $\le \delta \le $ 1.5, 0.1 $\le \eta \le $ 1.0 with interval of 0.1 mT, 0.1 and 0.1, respectively. 
In addition, fittings with 
more fine grid were done at a parameter region of -0.2 mT $\le B_c \le $ 0.2 mT, 
0.86 $\le \delta \le $ 1.10, 0.30 $\le \eta \le $ 0.66 with interval of 0.02 mT, 0.02 and 0.02, respectively. 
We created scaling plots of totally $\sim 1.2\times 10^4$ combinations of parameters. 
Each of them was fitted to Eq.~(\ref{scalingfunc}) and the correlation $R$ was evaluated. 
Here $\Lambda$, $A$ and $\alpha$ are free fitting parameters. 

\begin{figure}[tb]
\begin{center}
\includegraphics[width = 8.0 cm,clip]{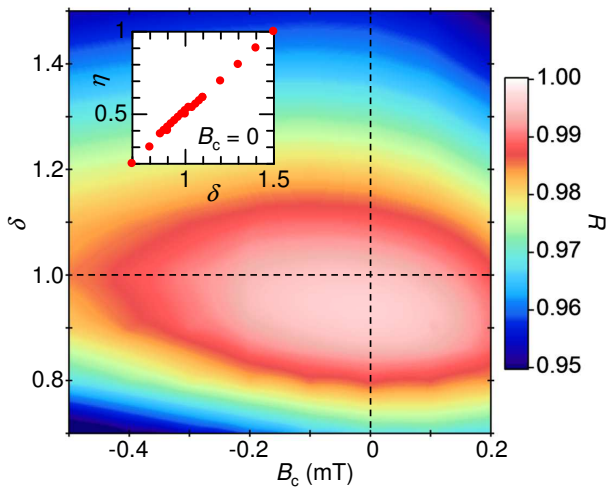}
\end{center}
\caption{
Contour plot of the correlation $R$ obtained from the fitting $\phi (x)$, where  $x=T/(B-B_{\rm c})^{\delta}$, 
to the scaling plot (see text).  
Here, $\eta$ is chosen for each $\delta$ with fixed $B_{\rm c}$ 
so that each $\eta$ gives maximum $R$ as shown in the inset as an example at $B_{\rm c}=0$. 
 } 
\label{Fig_scaling-quality}
\end{figure}

The best fit with the maximum value of $R=0.996$ is obtained at 
($B_{\rm c}$, $\delta$, $\eta$) = (-$0.02\pm 0.20$ mT, $0.94\pm 0.12$, $0.46\pm 0.12$). 
This result is consistent with (0, 1.0, 0.5) proposed originally in Ref.~\cite{matsumoto-ZFQCP}
, and further 
indicates that the data in the widest $T$ and $B$ range are scaled with this parameter set. 
The errors for the parameters $B_{\rm c}$, $\delta$ and $\eta$ were determined by the parameter space where $R\gsim 0.993$. 
Here we assumed that the error in $R$ is given by the difference between the maximum $R = 0.996$ and the ideal value of 1. 
The contour plot of $R$ is shown in Fig. \ref{Fig_scaling-quality}. 
Here, $\eta$ was chosen for each set of fixed $B_{\rm c}$ and $\delta$ 
so that it gives maximum $R$. An example at $B_{\rm c}$ = 0 is shown in the inset of Fig. \ref{Fig_scaling-quality}. 
This corresponds to adjusting $\eta$ against $\delta$ under fixed $B_{\rm c}$ so that a scaling is realized at high $T$ parts where the NFL power-law behavior is observed. 
The fitting results to the plots with ($B_{\rm c}$, $\delta$, $\eta$) = (0, 1.5, 1.0) and (0, 0.7, 0.2) 
are shown in Figs. \ref{Fig_scaling} (b) and (c) by the solid lines. 
These parameter sets are located at the top and bottom of the vertical broken line of $B_{\rm c}=0$ in Fig. \ref{Fig_scaling-quality} where 
$R=0.956$, 0.969, respectively. 
As is clearly seen in Fig. \ref{Fig_scaling-quality}, no other scaling is 
possible with the parameter set away from ($B_{\rm c}$, $\delta$, $\eta$) = (0, 1.0, 0.5).

From the above discussion, we confirm that the scaling law reported in Ref.~\cite{matsumoto-ZFQCP} 
provides the best fitting quality.
This sets an important bound for theoretical explanation of the QC behavior found in \ybal. 
First of all, this type of scaling behavior cannot be explained by the standard theory based on spin-density-wave type fluctuations\cite{matsumoto-ZFQCP}.
On the other hand, a recent study of 
the Anderson impurity model 
demonstrates that Kondo destruction at a mixed valent QCP is possible and indeed the $T/B$ scaling 
(and  $\omega / T$ scaling) is reproduced around the QCP \cite{pixley-PRL109}. 
It is an interesting future issue whether this is still relevant even in the lattice limit. 

Another interesting possibility has been discussed based on the anisotropy in the hybridization between the conduction and $f$ electrons ($c$-$f$ hybridization)\cite{ramires-PRL109}. 
From the argument based on the 
local symmetry of the Yb site, the crystal field ground doublet of both \aybal and \ybal is suggested to be made solely of $|J_z = \pm 5/2>$~\cite{Andriy09}. 
In this case, the $c$-$f$ hybridization is expected to be highly anisotropic 
as it has been already suggested experimentally for \aybal \cite{matsumoto-PRB84} 
and have a node along the $c$-axis. 
The theory proposes that this results in a heavy flat band 
with a $k^4$ dispersion and the QC in \ybal can be understood as a novel type of topological Fermi surface 
instability that should arise when the $f$ level is tuned to the bottom of the band. 
Interestingly, the theory reproduces the observed $T/B$ scaling in $M$ quite well.  
However, in this scenario, a pinning mechanism of the $f$ level to the bottom of the band remains as an open question. 
In order to examine the relevance of this theory, 
further studies, especially on the detail of the Fermi surface, are required.  

The effect of the critical valance fluctuations associated with a valence QCP or valence crossover \cite{watanabe-PRL105} 
might be also the key to understand the unconventional QCP in \ybal. 
Interestingly, a recent theory has demonstrated the robustness of the valence QCP against pressure 
to explain the QC found in an Yb-based quasicrystal \cite{watanabe-JPSJ82, deguchi-NatMater11}. 
Furthermore, it is worth noting that the theory reproduces the $T/B$ scaling especially at the NFL regime where $T/B \geq  0.8$ K/T\cite{watanabe-JPSJ83}. 
Measurements of the dynamical valence susceptibility will be the key to clarify whether the valence  fluctuations are the origin of the observed QC behaviors.

In any case,  there are apparently a lot of experimental and theoretical works to be done in order to 
understand the origin of the QC in \ybal. 
Among them, the mechanism of the HF formation even in the strong valence fluctuation may be the 
most fundamental problem to be answered. 
In the following, in order to give more insight into the HF behavior, we will discuss the details of the heavy Fermi-liquid component of $M$
obtained after subtracting the QC component, in particular focusing on its magnetic field evolution 
and the corresponding suppression of the HF behavior. 

\subsection{Heavy Fermi-liquid component of the magnetization}

\begin{figure}[tb]
\begin{center}
\includegraphics[width = 8.0 cm,clip]{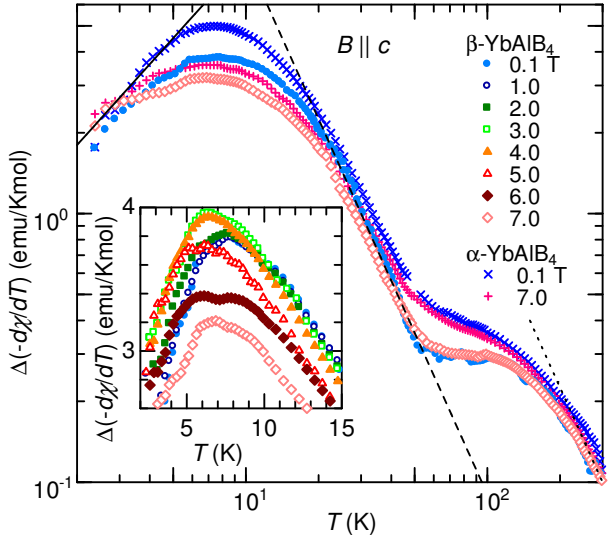}
\end{center}
\caption{
Heavy Fermi-liquid component in $-d\chi/dT$ of \ybal: $\Delta (-d\chi/dT)$ under the fields of 0.1 T and 7 T along the $c$-axis 
in the full logarithmic scale. 
Here, the QC component shown in Fig. \ref{Fig_scaling} (a) is subtracted.  
$-d\chi/dT$ of \aybal under the same fields (where no subtraction has been made) are also shown for comparison.  
The solid line is the guide to the eye which represents the $T$-linear behavior expected for a FL phase. 
The broken and dotted lines are also the guides to the eye which represent $T^{-2}$ dependence of  
the Curie(-Weiss) behavior in the high temperature limit (see text). 
Inset shows the field dependence of the peak found at $T\sim 8$ K for \ybal in a linear scale.  
 } 
\label{Fig_dMdT_deviation}
\end{figure}

As already discussed above,  while the QC component is suppressed by applying $B$, 
there is another component peaking at $T^*\sim$ 8 K in $-dM/dT$. 
This becomes dominant with increasing $B$ and 
finally overlaps on top of the one for \aybal at $B=7.0$ T (Fig. \ref{Fig_MoverH-dMdT} (b)).
In order to discuss the field evolution of this heavy Fermi-liquid component in detail, 
we subtracted the QC component at each $B$ by using the obtained scaling equation 
with the fitting parameters ($B_{\rm c}$, $\delta$, $\eta$) = (0, 1.5, 1.0) shown in Fig. \ref{Fig_scaling} (a). 
The obtained heavy Fermi-liquid component divided by $B$, \ie,  $\Delta (-dM/dT)/B = \Delta (-d\chi/dT)$ is shown in Fig. \ref{Fig_dMdT_deviation}. 
Here $-d\chi/dT$ of \aybal, where no subtraction is made, is also shown for comparison. 
Interestingly, the heavy Fermi-liquid component in \ybal is quite similar to $-d\chi/dT$ 
in \aybal even at a low field of $B=0.1$ T. 
This indicates that the magnetization of \ybal can be divided into two parts, 
the QC and the heavy Fermi-liquid components, the latter of which is similar to the one in \aybal. 

After peaking at $T^*\sim 8$ K, $\Delta(-d\chi/dT)$ exhibits 
a $T^{-2}$ power law behavior on heating up to 50 K (broken line in Fig. \ref{Fig_dMdT_deviation}), 
consistent with the Curie-Weiss behavior of paramagnetic local moments  
surviving above $T^*\sim 8$ K in both compounds.  
The Curie-Weiss fit to the $T$-linear  inverse susceptibility 
at a somewhat lower $T$ range extending even below $T^*\sim 8$ K ($6 \lsim T \lsim 15$ K) gives 
antiferromagnetic Weiss temperatures $\Theta_{\rm W}= 29$, 25 K and Ising moments $I_{\rm z} =$ 1.4, 1.3 $\mu _{\rm B}$ 
for the $\alpha$ and $\beta$ phases, respectively\cite{matsumoto-ZFQCP, matsumoto-SCES2010}. 
On the other hand, if we fit the asymptotic $T^{-2}$ power law behavior at $20\leq T\leq 50$ K to the temperature derivative of the Curie-Weiss law, 
we obtain $\Theta_{\rm W}= 4\pm1$, $2\pm1$ K for the $\alpha$ and $\beta$ phases, respectively, 
and $I_{\rm z} =0.6\pm0.1$ $\mu _{\rm B}$ for both phases.  
The latter evaluations could be more reliable in the sense that it is free from the constant term $\chi _0$ 
that has rather large error. 
If we adopt the latter moment sizes, the Wilson ratio becomes as high as 120 for both compounds. 
On the other hand, the low-$T$ part of the peak at $T^*\sim 8$ K 
exhibits a $T$-linear behavior (solid line in Fig. \ref{Fig_dMdT_deviation}) which is consistent with the FL groung state. 
Note that this is not clear for \ybal in the lower $B$ below $\sim 3$ T due to the QC component. 

Another $T^{-2}$ power law behavior found at a high $T$ range above 200 K 
corresponds to the high $T$ Curie-Weiss behavior reported previously, 
which gives antiferromagnetic $\Theta_{\rm W} = 110 \pm 2$, $108\pm 5$ K and Ising moments $I_{\rm z} = 2.22$, 2.24 $\mu _{\rm B}$ 
for $\alpha$ and $\beta$ phases respectively \cite{matsumoto-ZFQCP,matsumoto-SCES2010}. 
This crossovers to the low $T$ Curie-Weiss behavior across $T\sim$ 50 K where the $T$ dependence of 
$\Delta(-d\chi/dT)$ of \ybal and $-d\chi/dT$ of
\aybal indicate inflection points. Below this temperature, $M$ in both compounds exhibits further increase on cooling, which 
can be regarded as the onset of the HF behavior. 
Note that the $T$ derivative of the in-plane resistivity, $d\rho _{ab} /dT$, indicates a shoulder-like anomaly  at $T\sim 50$ K\cite{matsumoto-PRB84}. 
Furthermore, 
recent Hall effect measurements revealed a large peak in the $T$ dependence of $R_H$ at $\sim 40$ K.  
Both suggest coherence developing among $f$ electrons\cite{eoin-PRL109}.  


\begin{figure}[tb]
\begin{center}
\includegraphics[width = 8.0 cm,clip]{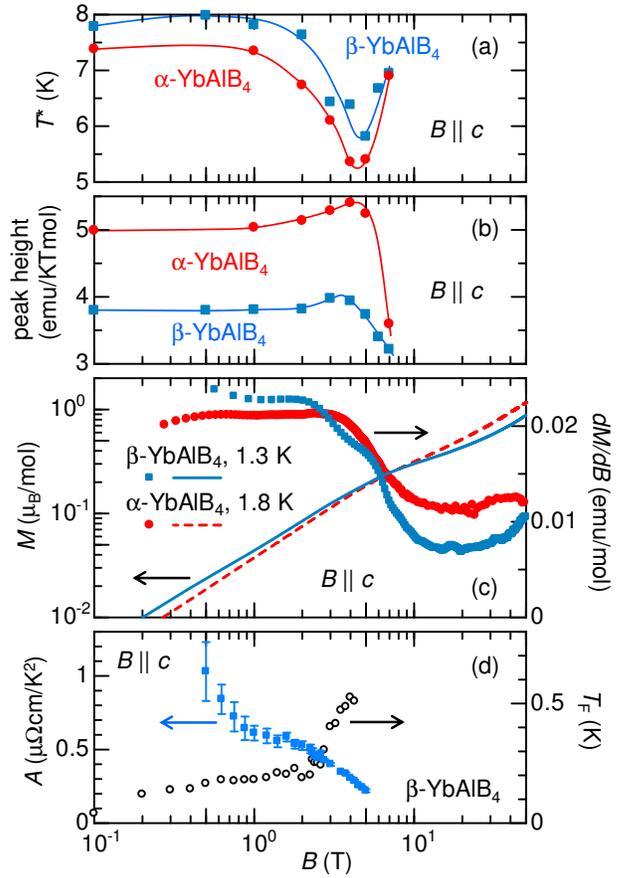}
\end{center}
\caption{
Field dependence of (a) the peak temperature $T^*$ and (b) its height for $\Delta(-d\chi/dT)$ of \ybal and $-d\chi/dT$ of
\aybal, respectively. 
(c) Magnetization curves (the left axis) and its field derivative $dM/dB$ (the right axis) in $\alpha$- and \ybal at $T = 1.8$ and 1.3 K, respectively. 
(d) Field dependence of the $A$ coefficient and $T_{\rm F}$ determined by the in-plane resistivity $\rho _{ab}$ of \ybal\cite{nakatsuji08}. 
The field is along the $c$-axis for all the figures. For details, see text.  
 } 
\label{Fig_B_dependence}
\end{figure}

\subsection{Suppression of the heavy fermion in $B$}
Next, we discuss the field evolution of the HF behavior in detail. 
As shown in the inset of Fig. \ref{Fig_dMdT_deviation}, 
$\Delta (-d\chi/dT)$ in \ybal changes its peak height and peak temperature $T^*$ with magnetic fields. 
The peak height slightly increases with field and maximizes at $B\sim$ 3 - 4 T and decreases rather steeply above this field. 
On the other hand, $T^*$ decreases by 30\% and shows the minimum at $B\sim$ 4 - 5 T. 
Figures \ref{Fig_B_dependence} (a) and (b) plot the $B$ dependence of $T^*$ 
and the peak height of both $\Delta (-d\chi/dT)$ in \ybal and $-d\chi/dT$ in \aybal, respectively. 
As clearly seen in the figure, their field evolutions are quite similar to each other 
except that $T^*$ in \ybal is larger by several \% than the one in \aybal and the peak height for \ybal is 30\% smaller than for \aybal. 
These field evolutions indicate a characteristic field scale of $B^*\sim 5$ T. 
Correspondingly, the magnetization curves for both compounds exhibit a slope change at $B^*$ (Fig. \ref{Fig_B_dependence} (c)). 

This field evolution can be regarded as the results of suppression of the HF behavior above $B^*\sim$ 5 T, 
as it was already discussed for \aybal \cite{matsumoto-SCES2013}. 
Furthermore, in the case of \aybal, 
it was discussed that the low $T$ part of the peak in $-d\chi/dT$, at least down to 2 K, deviates  
from FL behavior indicating a possible NFL behavior in the lower $T$ \cite{matsumoto-SCES2013}. 
Indeed, the recent transport and thermodynamic measurements down to very low temperatures below 0.1 K 
revealed an anisotropic NFL behavior which will be discussed elsewhere \cite{eoin-pre}. 
However, in \ybal, it is unlikely to have such a field induced NFL state at $B\sim$ 3 - 5 T in addition to the one found at zero-field.  
There has been no indication of NFL behavior  in the measurements such as 
the in-plane resistivity $\rho_{ab}$ down to 40 - 50 mK\cite{nakatsuji08} 
and specific heat at least down to $T=0.4$ K \cite{nakatsuji08, matsumoto-ZFQCP}. 
Furthermore, $-d\chi/dT$ at $B=3.0$, 4.0 T do not indicate clear NFL behavior (not shown). 
They well overlap to the one at 2.0 T indicating $T$-liniear dependence 
consistent with FL (Fig. \ref{Fig_MoverH-dMdT}(b)). 


The suppression of the HF behavior is also observed in the temperature dependence of $\rho_{ab}$.  
We present the $B$ dependence of $T_{\rm FL}$ and  $A$ coefficient determined by $\rho_{ab}$ in the previous work\cite{nakatsuji08}  
in  Fig. \ref{Fig_B_dependence} (d). Here, $T_{\rm FL}$ is defined as a temperature below which the resistivity exhibits the $T^2$ dependence.  
$A$ is the coefficient for the $T^2$ dependence, \ie, $\rho_{ab} = \rho_{ab, 0} + AT^2$ where $\rho_{ab, 0}$ is the in-plane residual resistivity. 
As already mentioned above, $T_{\rm FL}$ and $A$ are well-defined for all the field above 0.1 T 
suggesting that there is no field induced QC. 
On the other hand, both $T_{\rm FL}$ and $A$ indicate anomalies at $B\sim 2.3$ T which is somewhat smaller than the 
characteristic field scale of $B^*\sim 5$ T found in $M$. 
While $T_{\rm FL}$ suddenly increases above that field, $A$ decreases making a shoulder like feature at the same field. 
Both of these observations indicate suppression of the HF state.  
Note that the further increase of $A$ in the lower $B$ below 1 T corresponds to the zero field QC \cite{nakatsuji08}. 

In general, the HF behavior is expected to be suppressed when the Zeeman energy exceeds the Kondo coupling under magnetic fields. 
This is sometimes accompanied by a sudden localization of $f$ electrons (Kondo breakdown) or Lifsitz transition 
as suggested both experimentally and theoretically\cite{Bercx-PRB86}. 
In the case of $\alpha$- and \ybal,     
the suppression of the HF behavior at $B^*\sim5$ T is supposed to reflect the renormalized small Kondo scale.
If we assume that the Zeeman energy of the effective moment $g\mu _{\rm B}$ becomes comparable with the Kondo scale given by  $k_{\rm B}T^*$ 
at $B^*$, 
we obtain $g\mu _{\rm B}\sim$ 2.3 $\mu _{\rm B}$ by using $T^*=7.8$ K for \ybal at $B=0.1$ T. 
Interestingly, this is close to the value estimated from the $T/B$ scaling and its free energy, 1.94 $\mu _{\rm B}$\cite{matsumoto-ZFQCP}.
For \aybal, a similar value of $g\mu _{\rm B}\sim$ 2.2 $\mu _{\rm B}$ is obtained from $T^*=7.4$ K at $B=0.1$ T and $B^*=5$ T. 

Naively speaking, one would expect that a considerable amount of $f$ moments become polarized if a HF state is suppressed under magnetic field.  
However, in $\alpha$- and \ybal, 
$M$ only reaches $\sim 0.2$ $\mu _{\rm B}$ at $B^*$ 
(Fig. \ref{Fig_B_dependence} (c)) which is considerably smaller than the above estimation. 
This may indicate that only a fraction of $f$ moments participates in the HF behavior or QC. 
Interestingly, the recent Hall effect measurements in \ybal revealed that there are two bands having different Kondo temperatures $T_K$, namely, 
a hole like band with the coherence $T$ scale of 200 K and an electron like band with low carrier density ($\sim$10\%)
with the low $T_K$ of $\sim 40$ K. Indeed, this indicates that only the latter component 
is responsible for the HF behavior and QC in \ybal\cite{eoin-PRL109}.  
The effective moment of 0.6 $\mu _{\rm B}$ estimated from the low $T$ Curie-Weiss behavior 
corresponds to 30\% of the high $T$ value, suggesting again that only a fraction contributes to the HF and QC components. 

The suppression of the HF behavior with the similar anomaly in the magnetization curve has been observed in \yrs at 
$B\sim$ 10 T\cite{tokiwa-PRL94}. 
Interestingly, while thermodynamic measurements\cite{tokiwa-PRL94, gegenwart-NJP8, tokiwa-JMMM272-276} 
and the de Haas-van Alphen measurements\cite{rourke-PRL101} indicate only one characteristic field,  
recent thermal and electrical transport measurements have revealed that 
it consists of three successive Lifshitz transitions which interplay with a smooth suppression of the Kondo effect\cite{pfau-PRL110}.  
This may be also the case in \ybal where we found two slightly different field scales, $\sim 2.3$ and $\sim$ 5 T, in $\rho_{ab}$ and $M$, respectively. 
Further experimental studies, especially on thermal transport, will be important to clarify the origin of the multiple field scales. 
Note that, in \yrs, $M$ at the suppression field $\sim$ 10 T amounts to $\sim$ 1.1 $\mu _{\rm B}$. 
This is close to the value of 1.4 $\mu _{\rm B}$ estimated from 
low $T$ Curie-Weiss behavior\cite{Gegenwart02} indicating that the major part of the $f$-electrons becomes polarized. 
This is clearly different from our observations made in $\alpha$- and \ybal.  

We have not yet clarified the mechanism of the unusual HF formation under the strong valence fluctuations. 
However, the fact that only a fraction of $f$ moments participates in the HF behavior at least gives us a hint  
that the multiple bands and/or their topological feature may play an important role. 
Indeed, the two distinct $T$ scales observed in both compounds may be ascribable to 
the two Fermi surfaces having different Kondo temperatures due to the anisotropic $c$-$f$ hybridization. 
In this sense, studies on Fermiology are important future issues. 
On the other hand, it is an interesting  and challenging future issues to check and detect 
the strong valence fluctuation effect such as anomalous charge fluctuations associated with 
the valence criticality.  

\section{Conclusion}
We discussed the detailed evolution of  the magnetization ($M$) in \ybal with temperature and magnetic field.  
We presented the temperature dependence of $-dM/dT$ 
obtained over a four decades of $T$ range between 0.02 K and 320 K, which is wider than the previous report \cite{matsumoto-ZFQCP}. 
First, we checked the $T/B$ scaling found at the low $T$ and  the low $B$ region in detail.
In particular, we examined the possibility of other scaling such as $T/(B-B_{\rm c})^{\delta}$ scaling with $B_c\neq 0$ or $\delta\neq 1$, 
and confirmed that the $T/B$ scaling yields the best quality for the fitting. 

We further estimated the heavy Fermi-liquid component of $M$ after subtracting the QC component by using the $T/B$ scaling. 
The obtained heavy Fermi-liquid component exhibits the $T$ and $B$ dependence quite similar to the magnetization of 
\aybal, having a peak in $-dM/dT$ at $T^*\sim 8$ K which 
is suppressed above $B^*\sim 5$ T. 
This indicates that the HF behavior in $\alpha$- and \ybal becomes suppressed above this field corresponding to
a small renormalized Kondo scale. 
This suppression of the HF behavior may be related to 
the recent observation of the field induced anisotropic NFL behavior in \aybal\cite{eoin-pre}. 
In \ybal, on the other hand, we found no sign of such field induced NFL behavior.   
We note that it is an interesting question whether and how the two components found in the intermediate valence material are related to the two-fluid description 
of the Kondo lattice \cite{nakatsuji-PRL92}.

While the effective $g$-factor estimated from the Zeeman coupling relation $k_{\rm B} T^* = g \mu_{\rm B} B^*$ and $T/B$ scaling suggests the 
the effective moment of $\sim$ 2 $\mu _{\rm B}$ for \ybal, 
$M$ at the suppression field $B^*$ only reaches a considerably smaller value $\sim$ 0.2 $\mu _{\rm B}$.
This indicates that only a part of the $f$ moments participates in the zero-field QC or HF behavior, 
as also suggested by the 
recent Hall effect measurements\cite{eoin-PRL109}. 
The multiple bands and their topological feature should play an important role in 
the formation of the low temperature heavy fermion state and quantum criticality in the intermediate valence state.

\begin{acknowledgments}
We thank E. C. T. O'Farrell, P. Coleman, A. H. Nevidomskyy, 
S. Watanabe, C. Broholm, K. Ueda 
for supports and 
useful discussions.
This work is partially supported by Grants-in-Aid (No.
21684019) from JSPS, by
Grants-in-Aids for Scientific Research on Innovative Areas (No. 20102007, No. 21102507) from MEXT, Japan.
\end{acknowledgments}

%

\end{document}